\documentclass{iaus}
\usepackage{graphicx}

\title[short title of paper] 
{Searching for RR Lyrae stars in the Canis Major overdensity}

\author[C. Mateu, K. Vivas, R. Zinn \and L. Miller]   
{C. Mateu$^{1,2}$, K. Vivas$^1$, R. Zinn$^3$ \and L. Miller$^3$}

\affiliation{$^1$Centro de Investigaciones de Astronom\'{\i}a (CIDA), M\'erida, Venezuela\\
$^2$Facultad de Ciencias, Universidad Central de Venezuela, Caracas, Venezuela \\ 
$^3$Astronomy Department, Yale University, New Haven, USA.
}
\pubyear{2004}
\volume{241}  
\pagerange{119--126}
\date{?? and in revised form ??}
\jname{Proceedings Title IAU Symposium}
\editors{A.C. Editor, B.D. Editor \& C.E. Editor, eds.}
\begin{document}

\maketitle

\begin{abstract}


The Canis Major overdensity (CMa) was initially proposed to be the remnant
of a tidally disrupting dSph galaxy. Since its nature is still subject of debate, the goal of 
the present work was to conduct a large-scale RR Lyrae survey in CMa, in order to see
if there is an overdensity of these stars. The survey spans a total
area of $\sim34$ sq. deg. with observations in V and R filters, made
with the 1.0m J\"urgen Stock Schmidt telescope at the National
Astronomical Observatory of Venezuela. Current results in a subregion, including
spectroscopic observations, show that the small number of RR Lyrae stars
found can be accounted for by the halo and thick disk components of our Galaxy.

\keywords{Galaxy: structure, Galaxy: stellar content, stars: variables: RR Lyrae}
\end{abstract}

\firstsection 
\section{Introduction}

The nature of the so-called Canis Major (CMa) overdensity has been the center of a lively debate ever since its discovery (Martin et al. 2004, Momany et al. 2004). It was originally proposed to be the remant core of a tidally disrupting dSph galaxy, and even though further observational evidence gathered in CMa has supported this view, it also has opened other possibilities for explaining its nature without invoking an origin external to our Galaxy. RR Lyrae (RRL) stars were selected as tracers because they are relatively easily discovered by their variability, are excellent standard candles which enables the determination of accurate distances, and are found in every dwarf satellite galaxy of the Milky Way that has been adequately searched (Vivas and Zinn 2006). The presence or absence of an overdensity of RRLs in the CMa region may then help resolve the nature of this feature. 

\section{Color-Magnitude Diagram and RR Lyrae Star Search}

Multi-epoch observations in V and R filters were made with the $8$k x $8$k QUEST camera at the 1m J\"urgen Stock Telescope at the Venezuelan National Observatory. The survey covers a total area of $34$ sq. deg. Here we present the results of the RRL star search in a sub-region of $8.34$ sq. deg, and the color-magnitude diagram for a subregion of $17$ sq.deg. The saturation and limiting V magnitudes were 13 and 19.5 respectively. The color-magnitude diagram $V_o$ vs $(V-R)_o$ was obtained with psf photometry for a subregion spanning $17$ sq. deg. which include $\sim130000$ stars (Figure 1). No features such as a red clump, red giant branch or horizontal branch are obvious.

\begin{figure*}[!htb]
\begin{center}
  \includegraphics[width=5.1cm,height=5.1cm]{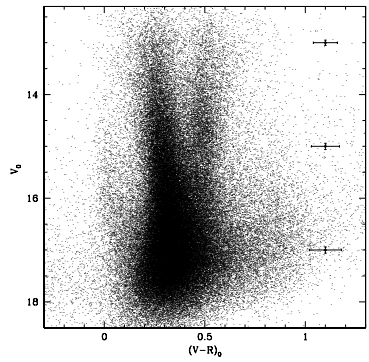}
  \caption{Color magnitude diagram for a 17 sq. deg. subregion in CMa.}
\end{center}
\end{figure*}

\section{RR Lyrae star search}\label{sec:rrl}



Variable stars were detected using a $\chi^2$  test (Vivas  et al. 2004). RRL star candidates were selected by constraining their $V-R$ colors to the expected range for RR Lyrae stars, and then fitting typical light curve templates (Layden 1998) to the data. The light curves were completed using additional photometric observations from the SMARTS telescopes at CTIO (Chile). Five strong candidates to be RRL stars were observed spectroscopically using the SMARTS 1.5m telescope. Three of these were confirmed as RRL stars using the spectroscopic data. Their radial velocities and metallicities are shown in Table 1.

\begin{table}[!h]
\begin{center}
\begin{tabular}{lccc}\hline
Star & $D_\odot$ (kpc) & $V_{rad}$ (km/s) & $[Fe/H]$ \\
\hline
882   & 6.2 & 50 & -1.20 \\
18056 & 5.0 & 69 & -1.0  \\
7701  & 5.1 & 180 & -1.26 \\\hline
\end{tabular}
\caption{Heliocentric distance, radial velocity and metallicities, for the confirmed RRLs.}
\end{center}
\end{table}

\section{Conclusions}

Although the search for RRL stars has not been completed in the whole area, we did not find a large overdensity of these stars in an area of $8.3$ sq. deg. The radial velocities and metallicities of the $3$ RRL stars in the region are consistent with them belonging to the known components of the Galaxy, halo and thick disk. In particular, star 7701 has the typical properties of the halo population. Stars 882 and 18056 have similar radial velocities which agree with the expectation for the thick disk population. It may be argued that those two stars with similar velocities may belong to a distinct population like a dSph galaxy, with mean [Fe/H]=-1.1. 
However, until finishing the whole survey that would increase our statistics, we are reluctant to give  
a definitive conclusion.

\end{document}